\pdfoutput=1
\documentclass[11pt]{article}
\usepackage{tablefootnote}
\usepackage{hyperref}
\usepackage{ftnxtra}

\usepackage{acl}
\usepackage{times}
\usepackage{latexsym}
\usepackage{booktabs}
\usepackage{graphicx}
\usepackage{multirow}
\usepackage{amsfonts}
\usepackage{amsmath}

\usepackage[T1]{fontenc}
\usepackage[utf8]{inputenc}

\usepackage{microtype}

\title{Synthesizing Human Gaze Feedback for Improved NLP Performance}
\everypar{\looseness=-1}
\author{Varun Khurana\thanks{\hspace{1em}Equal Contribution} \\
  Adobe, IIIT Delhi \\
  \texttt{\small{varun19124@iiitd.ac.in}} \\\And
  Yaman Kumar Singla\footnotemark[1]\\
  Adobe, IIIT Delhi, SUNY-Buffalo \\
  \texttt{\small{ykumar@adobe.com}} \\\AND
  Nora Hollenstein \\
  University of Copenhagen \\
  \texttt{\small{nora.hollenstein@hum.ku.dk}} \\\And
  Rajesh Kumar \\
  Bucknell University \\
  \texttt{\small{rajesh.kumar@bucknell.edu}} \\\And
  Balaji Krishnamurthy \\
  Adobe \\
  \texttt{\small{kbalaji@adobe.com}} \\}

\begin{document}
\maketitle
% \vspace{-1in}
\begin{abstract}
Integrating human feedback in models can improve the performance of natural language processing (NLP) models. Feedback can be either explicit (\textit{e.g.} ranking used in training language models) or implicit (\textit{e.g.} using human cognitive signals in the form of eyetracking). Prior eye tracking and NLP research reveal that cognitive processes, such as human scanpaths, gleaned from human gaze patterns aid in the understanding and performance of NLP models. However, the collection of \textit{real} eyetracking data for NLP tasks is challenging due to the requirement of expensive and precise equipment coupled with privacy invasion issues. To address this challenge, we propose ScanTextGAN, a novel model for \textit{generating} human scanpaths over text. We show that ScanTextGAN-generated scanpaths can approximate meaningful cognitive signals in human gaze patterns. We include synthetically generated scanpaths in four popular NLP tasks spanning six different datasets as proof of concept and show that the models augmented with generated scanpaths improve the performance of all downstream NLP tasks.
\end{abstract}

%%%%%%%%%%%%%%%%%%%%%%%%%%%%%%%%
%%%%%%%%%%%%%%%%%%%%%%%%%%%%%%%%
\section{Introduction}

\begin{figure}[!t]
%\vspace*{-10mm}    \centering
    \includegraphics[width=1\columnwidth]{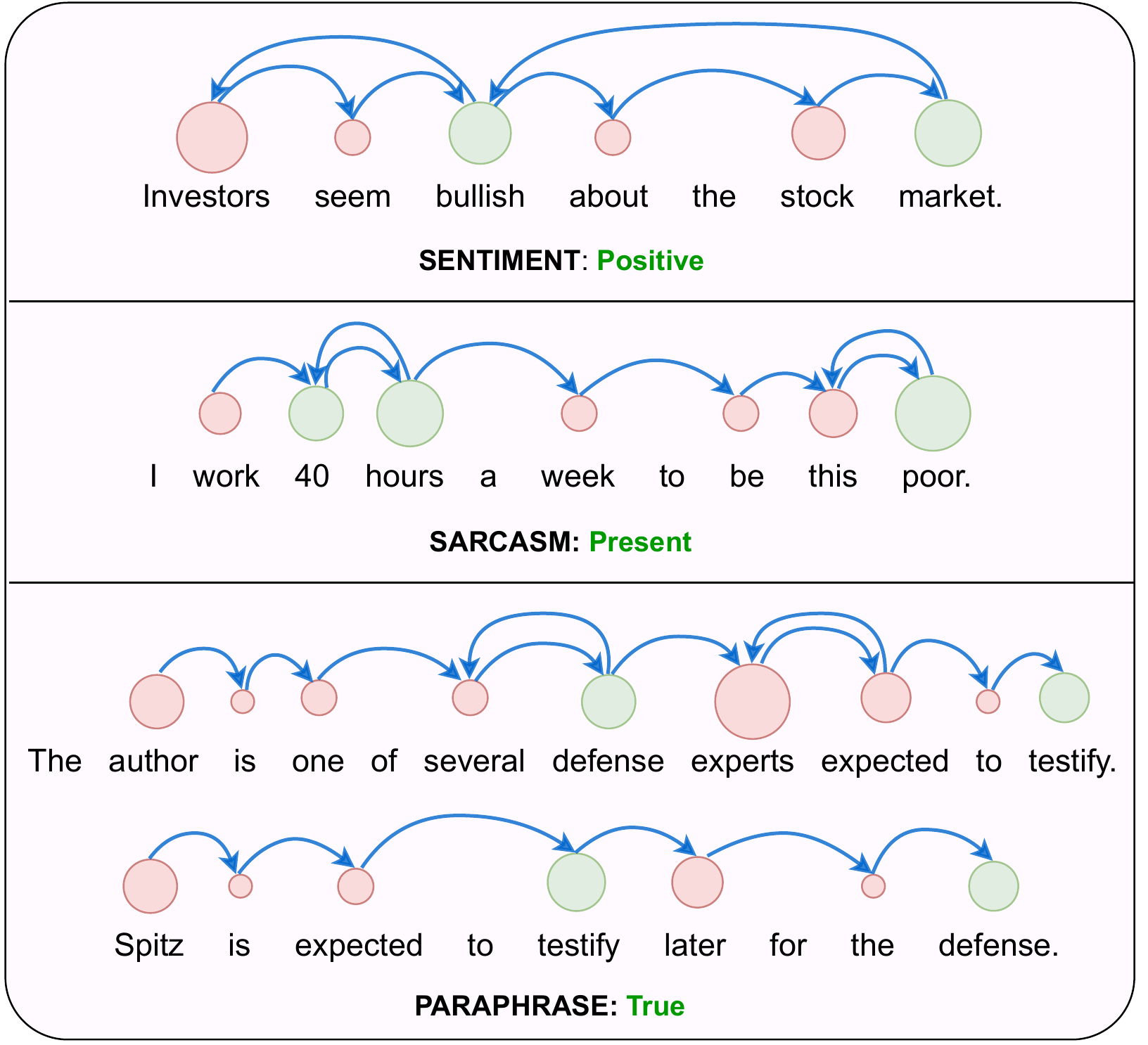}
    \caption{\small Generated scanpaths over text samples taken from various natural language processing (NLP) tasks. The green circles denote the important words characteristic of that task. The circles' size denotes the fixation duration, and the arrows depict the saccadic movements. As can be seen, linguistically important words often have a higher fixation duration and revisit. Regressions (word revisits) also appear in the examples. }
    \label{fig:scanpaths_for_NLP}
    \vspace*{-3mm}
\end{figure}

% \textbf{Eye tracking and scanpath}
% \textbf{NLP and real scanpath}
% \textbf{NLP and generated scanpath}
Integrating human signals with deep learning models has been beginning to catch up in the last few years. Digital traces of human cognitive processing can provide valuable signals for Natural Language Processing \cite{klerke2016improving,plank-2016-keystroke}. Various approaches for integrating human signals have been explored. For example, human feedback for better decisioning \citep{christiano2017deep}, NLP tasks \citep{stiennon2020learning,wu2021recursively}, and most recently language modeling using reinforcement learning with human feedback (RLHF) based reward \citep{bai2022training,ouyang2022training}. RLHF involves explicit human feedback and is expensive and hard to scale. On the other hand, previous studies have also tried to use implicit human feedback in the form of eyetracking signals.
%, human eye gaze for computer vision tasks like image captioning and visual question answering \citep{he2019human,boyd2022human}, and NLP tasks like sentiment analysis and NER \citep{NoraNEREyeTracking}.
It has proven to be a useful signal for inferring human cognitive processing \cite{sood2020improving, hollenstein-zhang-2019-entity, ijcaiSurveyGapIdentified}. NLP researchers have focused on assessing the value of gaze information extracted from large, mostly dis-jointly labeled gaze datasets in recurrent neural network models \cite{ren-xiong-2021-cogalign,strzyz-etal-2019-towards,barrett-etal-2018-sequence}. The proposed approaches under this paradigm include gaze as an auxiliary task in multi-task learning \cite{klerke-etal-2016-improving,hollenstein2019advancing}, as additional signals \cite{mishra-etal-2016-harnessing}, as word embeddings \cite{barrett-etal-2018-unsupervised}, as type dictionaries \cite{barrett-etal-2016-weakly,hollenstein-zhang-2019-entity}, and
as attention \cite{barrett-etal-2018-sequence}. 

Previous studies demonstrate that human scanpaths (temporal sequences of eye fixations, see Fig.~\ref{fig:scanpaths_for_NLP}) gleaned from eye tracking data improve the performance of NLP models. However, the real-world application of these methods remains limited primarily due to the cost of precise eye-tracking equipment, users' privacy concerns, and manual labor associated with such a setup. Therefore, generating scanpaths from existing eyetracking corpora would add great value to NLP research. To the best of our knowledge, this is the first paper to propose a model that generates scanpaths for a given read text with good accuracy. We call the model, ScanTextGAN.

\iffalse
    \begin{figure}[!t]
        %\vspace{-35mm}
        \centering
        \includegraphics[width=.9\columnwidth]{images/scanpath_sample_5.pdf}
        %\vspace{-8mm}
        \caption{\small A sample human scanpath, i.e., a temporal sequence of eye fixations and saccades over words in a sentence. The size of the circles denotes the fixation durations, and the arrows depict the saccadic movements. Regressions (word revisits) can also be recognized.}
        \label{fig:scanpath_sample}
        %\vspace{-5mm}
    \end{figure}
\fi

\begin{figure}[!t]
    % \vspace{-3mm}
    \centering
    \includegraphics[height=2in, width=0.97\columnwidth]{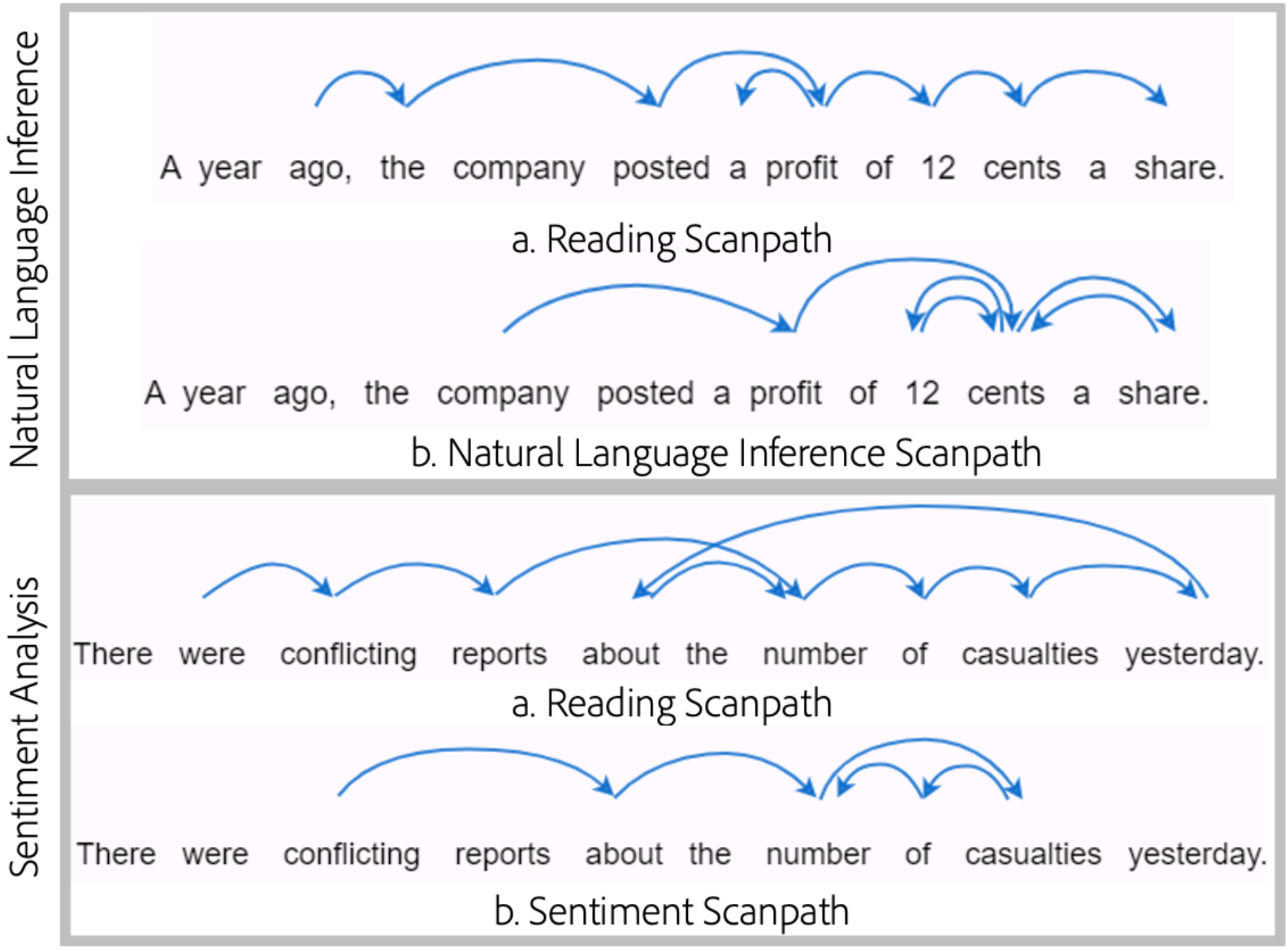}
    %\vspace*{-2mm}
    \caption{\small (Intent-aware) Scanpath samples generated by conditioning scanpath generation on different downstream natural language tasks. Note that the conditioned scanpaths are heavily biased to words important for that downstream task.}
    \label{fig:intent-scanpaths-example} 
%\vspace{-3mm}
\end{figure}

We demonstrate the scanpath generation capability of ScanTextGAN over three eye-tracking datasets using multiple evaluation metrics. Further, we evaluate the utility of \textit{generated} scanpaths for improvements in the performance of multiple NLP tasks (see Figs.~\ref{fig:scanpaths_for_NLP},\ref{fig:intent-scanpaths-example}) including the ones in the GLUE benchmark \cite{wang-etal-2018-glue}. The generated scanpaths achieve similar performance gains as the models trained with real scanpaths for classic NLP tasks like sentiment classification, paraphrase detection, entailment, and sarcasm detection. 

Our contributions are threefold:

\noindent \textbf{1.} We propose ScanTextGAN, the first scanpath generator over text. \\
\textbf{2.} We compare ScanTextGAN with multiple baselines and conduct ablation experiments with varying models and configurations. The model performs well on the test sets and cross-domain generalization on two additional eyetracking datasets belonging to different text domains.\\
\textbf{3.} We tested the usefulness of generated scanpaths in downstream NLP tasks such as sentiment analysis, paraphrase detection, and sarcasm detection on six different datasets. The results show that the downstream NLP tasks benefited significantly from cognitive signals inherent in generated scanpaths. Further, we show how scanpaths change when finetuning with downstream natural language tasks (Figs.\ref{fig:intent-scanpaths-example},\ref{fig:intent-saliency-example}) and that they lead to further improvements in downstream task performance (\S\ref{sec:intent-scanpaths}) showing how they can act as additional controls beyond the task architecture.

% On six different datasets, we utilize the generated scanpaths to model downstream NLP tasks such as sentiment analysis, paraphrase detection, and sarcasm detection and show improved performance due to the cognitive signals contained in generated scanpaths. The results demonstrate that our model yields well-corroborated predictions with the human gaze on out-of-domain data.

%The rest of the paper is organized as follows. Section \ref{sec:RelatedWork} discusses the related work and identified research gap. Section \ref{sec:ProposedModel} illustrates the proposed model and datasets used for training, the training process, and loss functions, followed by Section \ref{sec:Performance Evaluation}, which summarizes the evaluation and discusses the results. Finally, Section \ref{sec:ConclusionFutureWork} concludes the paper considering possible future work.  

%%%%%%%%%%%%%%%%%%%%%%%%%%%%%%%%
\section{Related Work}
\label{sec:RelatedWork}

When reading a text, humans do not focus on every word and often do not read sequentially \cite{Just1980}. A series of studies in psycho-linguistics have shown that the number of fixations and the fixation duration on a word depend on several linguistic factors. The linguistic factors can also be determined given the cognitive features \cite{clifton2007eye, demberg2008data}.
Though advances in ML architecture have helped bring machine comprehension closer to human performance, humans are still superior for most NLP tasks \cite{blohm-etal-2018-comparing,xia-etal-2019-automatic}. 

It has been shown in the literature that integrating explicit \citep{bai2022training,ouyang2022training} and implicit (cognitive processing) human feedback signals in traditional ML models is expected to improve their performance \cite{Just1980}. However, the cost of explicit feedback (e.g., using MTurk) and implicit feedback (e.g., eye tracking) at scale is excessively high. Similarly, privacy-invasive eye-tracking processes limit the scope of this idea. One way to address this problem is to use generated eye movements to unfold the full potential of eye-tracking research. Hence, the idea is to architect ScanTextGAN, a scanpath generator for text reading, and test its usefulness in downstream NLP tasks. 

% Integrating human cognitive processing signals in traditional ML models is expected to improve performance \cite{Just1980}. However, the major hindrances is the unavailability and unacceptability of expensive and privacy-invasive eye-tracking process. One way to address this problem is to use generated eye movements to unfold the full potential of eye-tracking research. This motivated us to build the first scanpath generator for text, ScanTextGAN, and show its performance over both eye-tracking datasets and in downstream NLP tasks. Our work builds upon previous works on 1)~human attention modeling and 2)~gaze integration in neural network architectures.
More precisely, this work builds upon previous works on 1)~human attention modeling and 2)~gaze integration in neural network architectures, which are described as follows:

\textbf{Human Attention Modeling:} Predicting what people visually attend to in images (saliency prediction) is a long-standing challenge
in neuroscience and computer vision, the fields have seen many data-based models \cite{wang2021salient}. In contrast to images, most attention models for eye movement behaviors during reading are cognitive process models, \textit{i.e.}, models that do not involve machine learning but implement cognitive theories \cite{engbert2005swift,xia-etal-2019-automatic}. Key challenges for such models are a limited number of parameters and hand-crafted rules. Thus, it is difficult to adapt them to different tasks and domains and use them as part of end-to-end trained ML architectures \cite{kotseruba202040}. In contrast, learning-based attention models for text remain under-explored. Within that, all eye tracking models are saliency prediction models with non-existent work in predicting scanpaths. On the other hand, visual scanpaths generation for image-based eye tracking data has been recently explored for both traditional \cite{PathGANScanPathGen} and 360$^{\circ}$ images \cite{ScanGAN360}.

\citet{matthies-sogaard-2013-blinkers} presented the first fixation prediction work for text. They built a person-independent model using a linear Conditional Random Fields (CRF) model. %A separate line of work has instead tried incorporating assumptions about the human reading process into the model design. For \textit{e.g.}, 
\citet{hahn-keller-2016-modeling} designed the Neural Attention Trade-off (NEAT) language model, which was trained with hard attention and assigned a cost to each fixation. Other approaches include sentence representation learning using surprisal and part of speech tags as proxies to human attention \cite{10.5555/3171837.3171864}.%, attention as a way to improve time complexity for NLP tasks \cite{seo2018neural}, and learning saliency scores by training for sentence comparison \cite{samardzhiev-etal-2018-learning}.

Our work differs from previous studies as we combine cognitive theory and data-driven approaches to predict scanpaths and further show its application in downstream NLP tasks \cite{hollenstein-etal-2021-multilingual,hollenstein-etal-2021-cmcl}.

\begin{figure*}[!t]
%\vspace*{-12mm}
    \centering
    \includegraphics[scale=0.54]{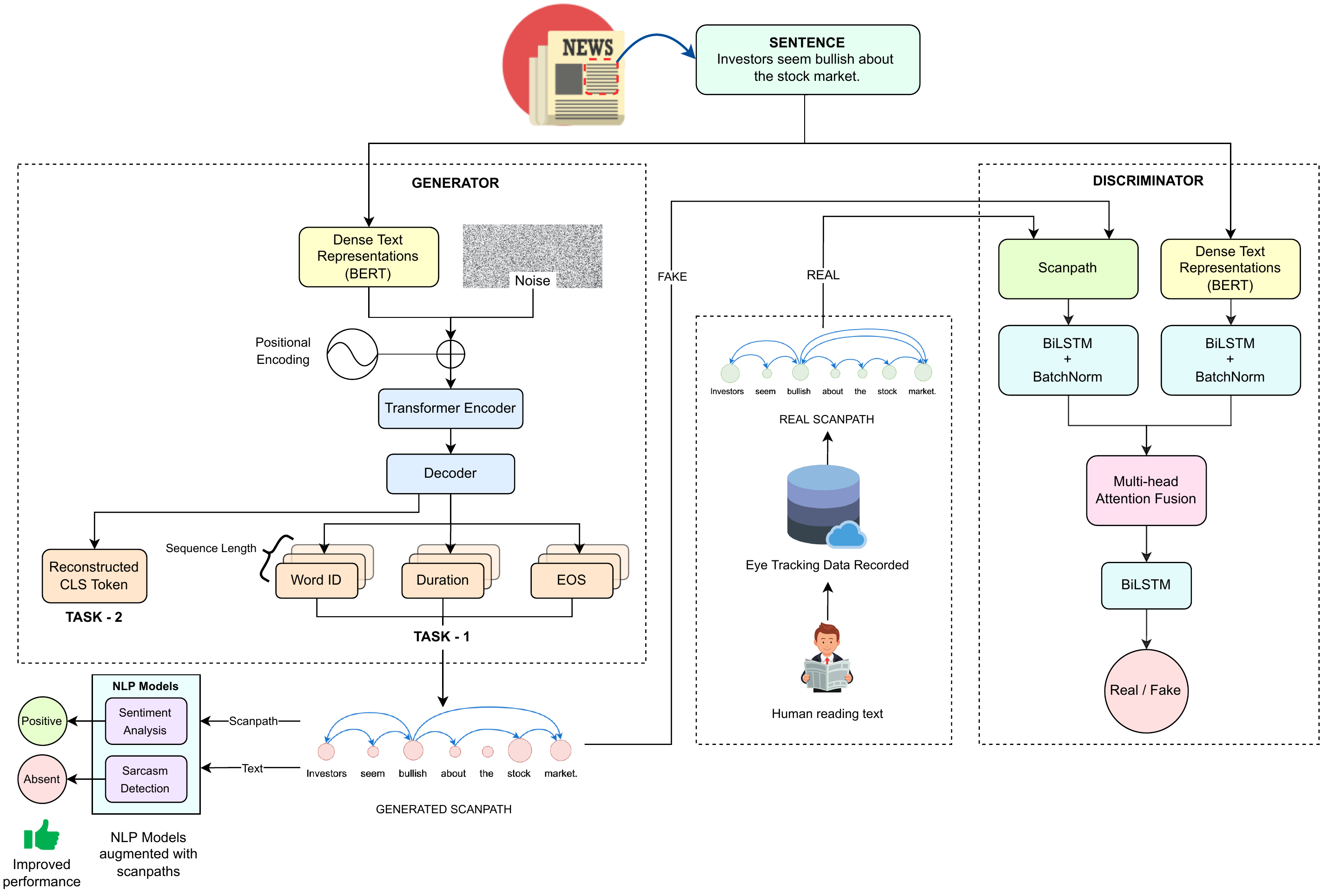}
    %\vspace*{-2mm}
    \caption{The architecture of the proposed \textbf{ScanTextGAN} model. The model consists of a conditional generator and a discriminator playing a zero-sum game. The generator is trained by two cognitively inspired losses: text content reconstruction and scanpath content reconstruction.}
    \label{fig:model} 
%\vspace*{-3mm}
\end{figure*}

\textbf{Integrating Gaze in Network Architecture:} Integration of human gaze data into neural network architectures has been explored for a range of computer vision tasks such as image captioning, visual question answering, and tagging \cite{karessli2017gaze,yu2017supervising,he2019human,boyd2022human}. %In language processing, tracking a reader's eye movements provides information about the cognitive processes of text comprehension \cite{RaynerReadingComp, Just1980}. 
Hence, recent research has utilized features gleaned from readers' eye movement to improve the performance of complex NLP tasks such as sentiment analysis \cite{long-etal-2017-cognition, mishra-etal-2016-leveraging}, sarcasm detection \cite{mishra-etal-2016-harnessing}, part-of-speech tagging \cite{barrett-etal-2016-cross}, NER \cite{hollenstein-zhang-2019-entity}, and text difficulty \cite{ScanPathApp1}.

While in recent years, eye tracking data has been used to improve and evaluate NLP models, the scope of related studies remains limited due to the %one of the main limitations of these methods of cognitively-inspired NLP is the %limited availability of large datasets and the 
requirement of real-time gaze data at inference time. \citet{ijcaiSurveyGapIdentified} reported that there exists no automated way of generating scanpaths yet in the literature. With high-quality artificially generated scanpaths, the potential of leveraging eyetracking data for NLP can be unfolded. Additionally, generating scanpaths that mimic human reading behavior will help advance our understanding of the cognitive processes behind language understanding. Hence, we propose ScanTextGAN; researchers can use that to generate scanpaths over any text without worrying about collecting them from real users. 

%We show the utility of the generated scanpaths by achieving performance gains in downstream NLP tasks over six corpora.

\section{Proposed Model}
\label{sec:ProposedModel}
In this section, we define the scanpath generation task, describe the ScanTextGAN model architecture, and provide details on loss functions and model training.

\textbf{Task Definition:} The task of scanpath generation is to generate a sequence $\mathcal{S}(\mathcal{T})$ representing a scanpath over the text $\mathcal{T} = \{w_1,w_2,...,w_n\}$ composed of a sequence of words, can be defined as follows:
\begin{equation}
    \mathcal{S(T)} = \{..,(w_a^i,t^i),....,(w_b^j,t^j),....,(w_c^k,t^k)\}
\end{equation}
where $t^i$ represents the fixation duration over the word $w_a$ occurring at the position $i$.  Note that it is not necessary to have $a<b$ (words being read in linear order) or that $k=n$ (the number of fixations being equal to the number of words). Due to regressions, \textit{i.e.}, backward saccades to previous words, words are also revisited. Hence, the same word could appear multiple times in the sequence.

\subsection{ScanTextGAN Model Architecture}
Fig.~\ref{fig:model} illustrates the proposed conditional GAN architecture of the model. The ScanTextGAN model is composed of two competing agents. First, a conditional generator that generates scanpaths given text prompts. The second is a discriminator network, which distinguishes real human scanpaths from the generated ones. The ScanTextGAN model is trained by combining text content loss, scanpath content loss, and adversarial loss (Eq.~\ref{eq:Net Generator Loss}). The scanpath content loss measures the difference between the predicted scanpath and the corresponding ground truth scanpath. The text content loss reconstructs the input text, and the adversarial loss depends on the real/synthetic prediction of the discriminator over the generated scanpath. We describe the losses along with the generator and discriminator architectures next.

\looseness=-1 \textbf{Generator:} The ScanTextGAN generator constitutes a transformer-based encoder-decoder framework. The encoder is conditioned on BERT-based text embeddings \cite{devlin-etal-2019-bert}, which are concatenated with noise to make the generator's output non-deterministic. The output of the transformer encoder is supplied to the decoder, which consists of task-specific feed-forward networks. One branch generates the scanpath (\textit{Task 1}), while the other reconstructs the $768$ dimensional CLS token embedding of the sentence (\textit{Task 2}). The scanpath is output as a temporal sequence of word ID (fixation points) $w_a^i$, fixation duration $t^i$, and end-of-sequence probability $EOS^i$. At inference time, the length $L(G)$ of generated scanpath $G$ is determined as follows:

\begin{equation}
L(G) = \begin{cases}
          \min_{1 \leq k \leq M} (k) \quad &\text{if} \, EOS^k > \tau \\
          M \quad &\text{otherwise} \, \\
     \end{cases}
\end{equation}
where $M$ is the maximum scanpath length as described in section \S\ref{sec:dataset} and $\tau \in (0,1)$ is a probability threshold. We use $\tau = 0.5$. The loss functions of the two branches are described below.

\looseness=-1 \textbf{Scanpath Content Loss} tries to minimize the deviation of generated scanpaths $\mathcal{G}(\mathcal{T}, \mathcal{N})$ from the ground-truth scanpaths $\mathcal{R}(\mathcal{T}, h))$ over text $\mathcal{T}$ where ground-truth scanpaths are recorded from the human $h$ and $\mathcal{N}$ stands for Gaussian noise $\mathcal{N}(0, 1)$. The loss function $\mathbb{L}_s$ is given as:
\begin{equation}
    \label{eq:Scanpath Content Loss}
    \begin{aligned}
    \mathbb{L}_s(\mathcal{G}(\mathcal{T,}\mathcal{N}), \mathcal{R}(\mathcal{T},h)) = \frac{1}{k} \Sigma_{i=0}^{k}(&\alpha(id_g^i-id_r^i)^2 + \\ \beta(t_g^i-t_r^i)^2 + &\gamma (E_g^i-E_r^i)^2)
    % \\+ &\gamma P_g(E^i)*log(P_r(E^i)))
    \end{aligned}
\end{equation}
which is a weighted sum of three terms. The first term measures the error between real and predicted \textit{fixation points} given by the mean squared difference between generated and real word-ids $(id_g^i-id_r^i)$. It penalizes permutations of word ids and trains the model to approximate the real sequence of fixation points closely.

The second term measures the difference in \textit{fixation durations} given by the mean squared difference between generated and real duration $(t_g^i-t_r^i)$. Fixation durations simulate human attention over words in the input text. Thus, a word with a larger fixation duration is typically synonymous with greater importance than other words in the input text. This error term supplements the generator's ability to learn human attention patterns over the input text.

Finally, the third term measures the mean squared error between the prediction of end-of-sequence probability by real and generated distributions $(E_g^i-E_r^i)$. These are weighted by the hyperparameters $\alpha,\beta$, and $\gamma$. Preliminary experiments showed that optimizing the mean squared error leads to better performance over the cross-entropy loss for optimizing the EOS probability output.

\looseness=-1 \textbf{Text Content Loss:} Scanpaths depend heavily on the linguistic properties of the input text. Therefore, to guide the generator towards near the probable real data manifolds, we adopt reconstruction of the CLS token embedding of the input text (\textit{Task 2}) by the generator as an auxiliary task since the CLS token embedding encodes a global representation of the input text. 
This text content reconstruction loss $\mathbb{L}_r$ is given as: 
\begin{equation}
    \label{eq:Text Content Loss}
    \begin{aligned}
     \mathbb{L}_r(\mathcal{G}(\mathcal{T,}\mathcal{N}), \mathcal{R}(\mathcal{T},h)) = (&BERT(w_i^g,w_j^g,...,w_k^g\\-&BERT(w_a^r,w_b^r,...w_n^r))^2
    \end{aligned}
\end{equation}
where $BERT(w_a^r,w_b^r,...w_n^r)$ and $BERT(w_i^g,w_j^g,...w_k^g)$ stand for the \textit{CLS} vector representations of real and generated text respectively.

\textbf{Discriminator:} The goal of the discriminator is to distinguish between the real and synthetic scanpaths supplied to it. Similar to the generator, it requires text representations to distinguish between real and generated scanpaths. Specifically, the discriminator comprises two blocks of BiLSTMs that perform sequential modeling over the scanpaths and BERT embeddings. The outputs of the two branches are combined and passed to an attention fusion module with four heads, followed by another network of BiLSTMs. The hidden states of the last BiLSTM layer from both forward and backward directions are concatenated and supplied to a feed-forward network. A Sigmoid function activates the output of the feed-forward network. In this manner, the discriminator classifies the input scanpaths as either \textit{real} or \textit{fake}.

\textbf{Adversarial Loss:} The generator and discriminator networks are trained in a two-player zero-sum game fashion. The loss is given by:
\begin{equation}
    \label{eq:Adversarial Loss}
    \begin{aligned}
    \mathbb{L}_a = \min_{G}\max_{D}\mathbb{E}_{x\sim p_{\text{data}}(x)}[\log{D(x|\mathcal{T},h)}] + \\  \mathbb{E}_{z\sim p_{\text{z}}(z)}[1 - \log{D(G(z|\mathcal{T,N}))}]
    \end{aligned}
\end{equation}
Therefore, the net generator loss becomes:
\begin{equation}
    \label{eq:Net Generator Loss}
    \begin{aligned}
    \mathbb{L}_g = \mathbb{L}_s + \mathbb{L}_r + \mathbb{E}_{z\sim p_{\text{z}}(z)}[1 - \log{D(G(z|\mathcal{T,N}))}]
    \end{aligned}
\end{equation}

\subsection{Dataset} 
\label{sec:dataset}
\looseness=-1 For training the ScanTextGAN model, we use the CELER dataset \cite{berzak2022celer}. It contains eyetracking data of 365 participants for nearly 28.5 thousand newswire sentences, sourced from the Wall Street Journal Penn Treebank \cite{marcinkiewicz1994building}. Each participant in CELER reads 156 newswire sentences. Half of the sentences are shared across participants, and the rest is unique to each participant. The maximum sentence length was set to 100 characters. Participant eyetracking data were recorded using Eyelink 1000 tracker in a desktop mount configuration with a sampling rate of 1000 Hz. The ScanTextGAN model is trained to approximate the average eye movements of all the participants who read given sentences. The CELER dataset was envisioned to enable research on language processing and acquisition and to facilitate interactions between psycholinguistics and natural language processing. Furthering the goal, we use it to train our conditional GAN model through which we show human scanpath approximation capabilities (\S\ref{sec:Evaluation of Scanpath Generation}). Also, we use it to show improvements in the performance of NLP tasks (\S\ref{sec:Application to NLP Tasks}). 

\looseness=-1
The data consist of tuples of participant ID, sentence ID, and word ID corresponding to fixation point and fixation duration. We compute the 99th percentile of fixation durations and treat it as the largest value. Fixations of durations longer than this are treated as outliers and hence dropped from the dataset. To apply the scanpath reconstruction loss (Eq.~\ref{eq:Scanpath Content Loss}), we scale all fixation durations by the maximum value and then normalize them to [0,1]. Similarly, word IDs in each sentence are normalized to [0, 1] after scaling them by the length of that sentence. For the last fixation point in every scanpath, the binary EOS token is set to 1. The maximum scanpath length is set to 80 fixation points (99th percentile of the lengths). Thus shorter scanpaths are padded while longer scanpaths are trimmed. We use BERT to encode the sentences and obtain their $768$-dimensional embeddings, keeping the max length parameter as 80, thus resulting in an $80\times768$ dimensional tensor.

\subsection{Parameter Settings}
\looseness=-1
Sinusoidal positional encoding is applied over the input embeddings fed to the generator. We use a 3-layer transformer encoder with four head attention and a hidden dimension size of 776 in the generator. In the discriminator, we use bidirectional LSTMs over sentence embeddings and generated scanpaths with a hidden size of 64 and a dropout ratio of 0.3, followed by batch normalization for faster convergence. An attention module with four attention heads is applied after concatenating the outputs.
We employ the Adam and RMSProp optimizer to minimize generator and discriminator losses. The batch size is set to 128, the initial learning rate of the generator to 0.0001, and that of the discriminator to 0.00001. The model is trained for 300 epochs. Our implementation uses PyTorch, a popular deep-learning framework in Python. All experiments are run on an Intel Xeon CPU with Nvidia A100-SXM GPUs.

\section{Performance Evaluation}
\label{sec:Performance Evaluation}
We quantify the performance of ScanTextGAN in two regimes\footnote{All results are calculated with five random seeds and reported as the mean of those five runs}; first, scanpath generation with three datasets, and second, NLP tasks with six datasets. Similar to prior computer vision studies \cite{sun2019visual,de2022scanpathnet,kummerer2021state,jiang2016learning}, we evaluate the ScanTextGAN model over the scanpath generation task. For this, we use the test split of the CELER dataset, \citet{mishra2016predicting}, and \citet{Mishra_Kanojia_Nagar_Dey_Bhattacharyya_2017}. In addition, unlike the computer vision studies, we also evaluate the ScanTextGAN model for improvement in NLP tasks. The hypothesis is that the human eyes (and consequently the brain) process many language comprehension tasks unconsciously and without visible effort. The next logical step is to capture (or, in our case, generate) this mental representation of language understanding and use it to improve our machine-learning systems. For evaluation, we use four tasks from the GLUE benchmark and two from the tasks proposed by \citet{mishra2016predicting}. While the ScanTextGAN model is trained over news text from the CELER dataset, with the help of the other datasets, we expand our testing to other domains, including reviews, quotes, tweets, and Wikipedia text.

\subsection{Evaluation Datasets}
\label{sec:eval_datasets}
\textbf{\citet{Mishra_Kanojia_Nagar_Dey_Bhattacharyya_2017}} comprises eye movements and reading difficulty data recorded for 32 paragraphs on 16 different topics, \textit{viz.} history, science, literature, \textit{etc}. For each topic, comparable paragraphs were extracted from Wikipedia\footnote{\url{https://en.wikipedia.org/}} and simple Wikipedia\footnote{\url{https://simple.wikipedia.org/}}. The participant's eye movements are tracked using an SR-Research Eyelink-1000 Plus eye tracker. Using the ground truth scanpaths over the text corpora, we evaluate the quality of generated scanpaths.

\textbf{\citet{mishra2016predicting}} contains eye fixation sequences of seven participants for 994 text snippets annotated for sentiment and sarcasm. These were taken from Amazon Movie Corpus %\cite{pang-lee-2004-sentimental}
, Twitter, and sarcastic quote websites. %They used an SR-Research Eyelink-1000 eye-tracker to collect the eye movements of the participants. 
The task assigned to the participants was to read one sentence at a time and annotate it with binary sentiment polarity labels (\textit{i.e.}, positive/negative). %Using feature engineering approaches, 
%They used the scanpath data to show improvements in sarcasm detection. 
The same datasets were used in several studies \cite{joshi-etal-2015-harnessing,mishra-etal-2016-harnessing,mishra-etal-2016-leveraging} to show improvements in sarcasm and sentiment analysis.
We use the datasets to evaluate both the generation quality and potential improvements in NLP tasks.% using generated scanpaths, thereby making it more aligned with real-world settings and solving the problem of the unavailability of human scanpath data at inference time.

\begin{table*}[!t]\centering
% \vspace*{-10mm}
% \scriptsize
\resizebox{\textwidth}{!}{\begin{tabular}{lcccccc}\toprule
\multirow{2}{*}{\textbf{Generator Model}} &\multicolumn{4}{c}{\textbf{MultiMatch $\uparrow$}} &\multirow{2}{*}{\textbf{Levenshtein Distance $\downarrow$}} \\\cmidrule{2-5}
&\textbf{Vector$\uparrow$} &\textbf{Length$\uparrow$} &\textbf{Position$\uparrow$} &\textbf{Duration$\uparrow$} & \\\midrule
Inter-subject score\footnotemark & 0.973   & 0.958 &    0.830   & 0.698 & 0.691\\\midrule
%Transformer Encoder-Decoder &0.974 &0.958 &\textbf{0.819} &0.543 & \\
%Transformer Encoder &0.974 &0.957 &0.773 &0.703 & \\
LSTM Encoder-Decoder trained with scanpath content loss & 0.975 & 0.956 &  0.765 & 0.344 & 0.865\\
ScanTextGAN -- Text Reconstruction -- GAN Loss& 0.968 & 0.947 & 0.728 & 0.703 & 0.779\\
\textbf{ScanTextGAN} &\textbf{0.983} &\textbf{0.972} & \textbf{0.787} & 0.733 & \textbf{0.769}\\
ScanTextGAN -- Text Reconstruction &0.974 &0.957 &0.773 &0.703 & 0.798\\
ScanTextGAN -- GAN Loss &0.973 &0.955 &0.750 &\textbf{0.761} &0.786\\
% *ScanTextGAN + Dwell Time reconstruction & 0.960 & 0.932 & \textbf{0.805} & 0.711 \\
% *Dwell Time reconstruction w/o Adv. training & 0.966 &0.939	&0.786 &0.775 \\
ScanTextGAN + addition of noise &0.971 &0.952 &0.756 &0.736 &0.791\\
ScanTextGAN -- Text (CLS) Reconstruction + sentence reconstruction &0.978 &0.963 &0.724 &0.721 &0.805 \\
% \midrule
% Inter-Subject Scanpath Topline & XXX & & & & 
% \\ 
\bottomrule
\end{tabular}}
\caption{In-domain Evaluation of Scanpath Generation on the CELER dataset \cite{berzak2022celer}.}\label{tab:celer}
%\vspace*{-3mm}
\end{table*}

\footnotetext{In the CELER dataset, there are only 78 shared sentences amongst all the participants. Therefore, inter-subject scanpath evaluation is done only for these sentences. In contrast, the ScanTextGAN results are reported for the entire test set (including these 78 sentences).}

Furthermore, we explore the potential of including cognitive signals contained in scanpaths in NLP models for a range of GLUE tasks which include Sentiment Analysis using Stanford Sentiment Treebank (SST), Paraphrase Detection using Microsoft Research Paraphrase Corpus (MRPC) and Quora Question Pairs (QQP), Natural Language Inference using Recognizing Textual Entailment (RTE) dataset.
%- \textbf{SST}: The Stanford Sentiment Treebank includes fine-grained sentiment labels for 215154 phrases in the parse trees of 11855 sentences. The corpus consists of sentences from movie reviews and human sentiment annotations. The task is to predict the sentiment of a given sentence. We use the two-way (positive/negative) class split and only sentence-level labels.\\
%- \textbf{MRPC}: The Microsoft Research Paraphrase Corpus comprises a set of 5,801 sentence pairs collected from newswire articles. Each pair is annotated with a label indicating whether the sentences are paraphrased or not.\\
% Each pair is labeled whether human annotators paraphrase it. The dataset is divided into \textit{train} (4,076 sentence pairs, of which 2,753 are paraphrases) and \textit{test} (1725 pairs, of which 1,147 are paraphrases).\\
%- \textbf{QQP}: The Quora Question Pairs (QQP) dataset consists of over 400,000 question pairs. Each is annotated with a binary value indicating whether the two questions are paraphrases of each other. \\
%- \textbf{RTE}: The Recognizing Textual Entailment (RTE) dataset involves a generic Natural Language Inference task that recognizes, given a pair of texts, whether the meaning of one text can be inferred from the other. 

Next, we cover the results of scanpath generation and its application in NLP tasks.

\begin{figure}[]
    \centering
    %\vspace{-3mm}
    \includegraphics[width=.85\columnwidth]{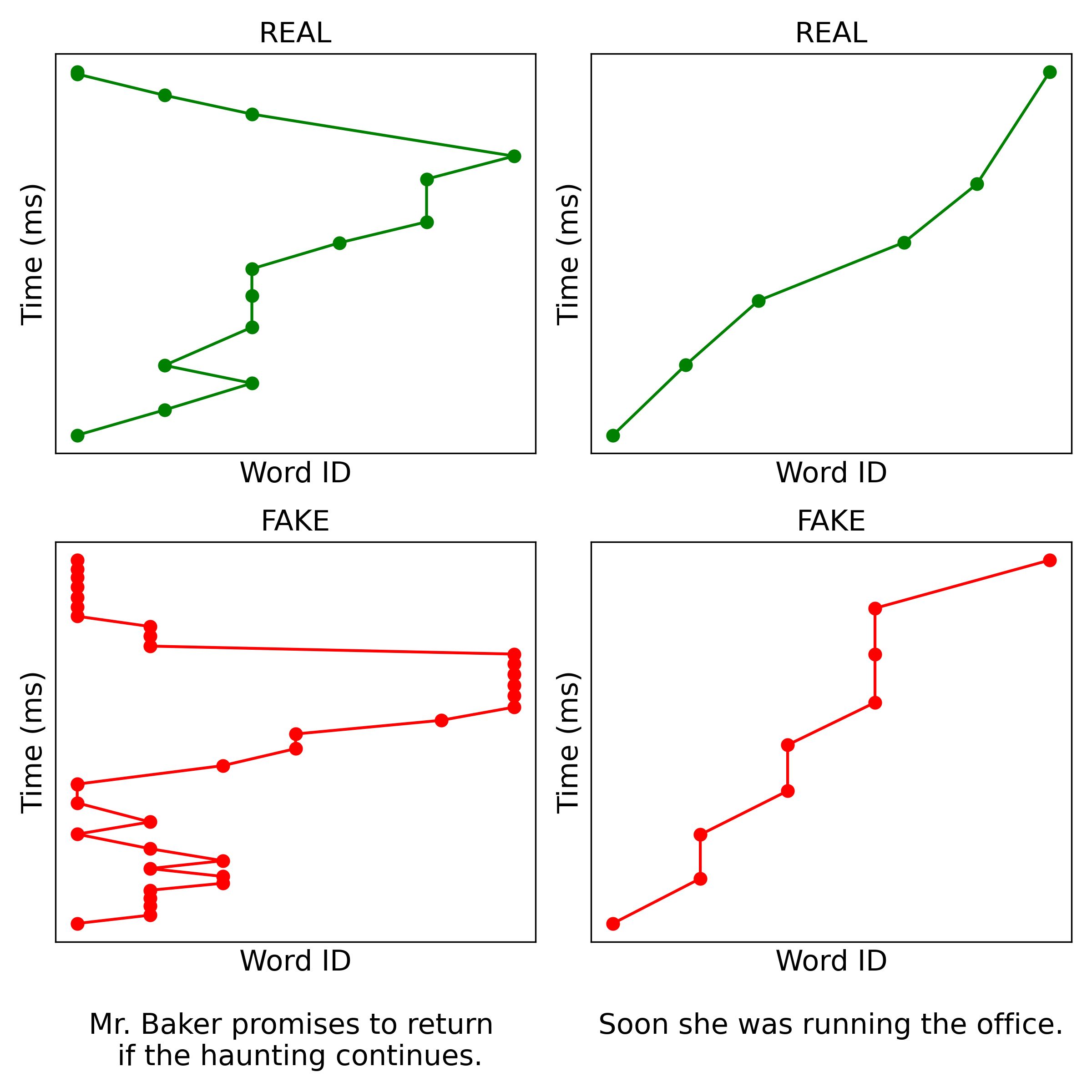}
   % \vspace{-2mm}
    \caption{Comparison of \textit{real} and \textit{synthesized} scanpaths corresponding to a few text samples. The proposed ScanTextGAN model generates the latter.}
    \label{fig:scanpath_comparison}
    %\vspace{-3mm}
\end{figure}

\subsection{Evaluation of Scanpath Generation}
\label{sec:Evaluation of Scanpath Generation}
We evaluate the scanpath generation model on two most commonly used metrics in image scanpath generation studies \cite{sun2019visual,chen2018scanpath,de2022scanpathnet,kummerer2022deepgaze}: \textbf{MultiMatch} \cite{jarodzka2010vector} and \textbf{Levenshtein Distance} \cite{levenshtein1965leveinshtein}. Multimatch is a geometrical measure that compares scanpaths across a comprehensive set of dimensions composed of shape, lengths, position, and fixation duration. Levenshtein Distance between a pair of sequences measures the least number of edits (inserts, deletes, substitution) to transform one into the other. More details are discussed in Appendix:\ref{sec:appendix_scanpath_metrics}.

Further, as a top-line comparison, we use \textbf{inter-subject scanpath similarity} \cite{sun2019visual}. It measures the degree of variation among real human scanpaths corresponding to each text input. To compute this, we first calculate each subject's performance by treating the scanpaths of other subjects as the ground truth. Then, the average value of all subjects is used as inter-subject performance.

\textbf{Baselines:} Since ScanTextGAN is the first text-based scanpath generation model, we conduct an ablation study to compare ScanTextGAN with its other variants. Specifically, we compare ScanTextGAN with the following six configurations: (1)~An LSTM-based network trained with scanpath content loss. Sentence embeddings obtained through BERT are concatenated with noise in this model. The resultant is fed to an attention module with four heads, then passed to a network of LSTMs and Batch Normalization layers applied in tandem. (2)~ScanTextGAN model trained with only the scanpath content loss. (3)~ScanTextGAN model without the text reconstruction loss (Task-2). (4)~ScanTextGAN model with BERT-based sentence embeddings reconstruction instead of CLS token reconstruction. (5)~ScanTextGAN model with the addition of noise instead of concatenation. (6)~ScanTextGAN model trained without GAN loss.

\begin{table*}[!t]\centering
%\vspace*{-10mm}
% \scriptsize
\resizebox{0.9\textwidth}{!}{\begin{tabular}{lcccccc}\toprule
\multirow{2}{*}{\textbf{Generator Model}} &\multicolumn{4}{c}{\textbf{MultiMatch $\uparrow$}} &\multirow{2}{*}{\textbf{Levenshtein Distance $\downarrow$}} \\\cmidrule{2-5}
&\textbf{Vector$\uparrow$} &\textbf{Length$\uparrow$} &\textbf{Position$\uparrow$} &\textbf{Duration$\uparrow$} & \\\midrule
Inter-subject score & 0.977   & 0.963 &    0.839   & 0.715 & 0.723\\\midrule
LSTM Encoder-Decoder trained with scanpath content loss& \textbf{0.984}    & \textbf{0.973} &   0.714 &   0.379 & 0.918\\
ScanTextGAN -- Text Reconstruction -- GAN Loss& 0.977   & 0.960 &    0.780   & 0.769 & 0.847\\
%Transformer Encoder-Decoder &0.977 &0.957 &0.821 &0.574 & \\
%Transformer Encoder &0.976 &0.961 &0.763 &0.757 & \\
\textbf{ScanTextGAN} &0.966 &0.945 &\textbf{0.791} &\textbf{0.771} & \textbf{0.836} \\
ScanTextGAN -- Text Reconstruction &0.976 &0.961 &0.763 &0.757 & 0.845\\
% *ScanTextGAN + Dwell Time reconstruction & 0.971 &0.952 & \textbf{0.797} &0.711 \\
% *Dwell Time reconstruction w/o Adv. training & 0.976 &0.959	&0.795 &0.765 \\
ScanTextGAN -- GAN Loss &0.976 &0.959 &0.774 &0.768 &0.839\\
ScanTextGAN + addition of noise &0.968 &0.947 &0.737 &0.743 & 0.838\\
ScanTextGAN -- Text (CLS) Reconstruction + sentence reconstruction &0.964 &0.934 &0.747 &0.733 & 0.869\\
\bottomrule
\end{tabular}}
\caption{Cross-domain Evaluation of Scanpath Generation on the Dataset by \citet{mishra2016predicting}.}\label{tab:iitb_2016}
%\vspace*{-3mm}
\end{table*}

\begin{table*}[!t]\centering
% \vspace*{-10mm}
% \scriptsize
\resizebox{0.9\textwidth}{!}{\begin{tabular}{lcccccc}\toprule
\multirow{2}{*}{\textbf{Generator Model}} &\multicolumn{4}{c}{\textbf{MultiMatch $\uparrow$}} &\multirow{2}{*}{\textbf{Levenshtein Distance $\downarrow$}} \\\cmidrule{2-5}
&\textbf{Vector$\uparrow$} &\textbf{Length$\uparrow$} &\textbf{Position$\uparrow$} &\textbf{Duration$\uparrow$} & \\\midrule
Inter-subject score & 0.994   & 0.991 &    0.834   & 0.620 & 0.845\\\midrule
LSTM Encoder-Decoder trained with scanpath content loss& \textbf{0.992}  &  \textbf{0.987}   & 0.596 &    0.329 & 0.969\\
ScanTextGAN -- Text Reconstruction -- GAN Loss&0.990 &0.984 &0.729 &0.705 & 0.951\\
%Transformer Encoder &0.986 &0.981 &0.776 &0.706 & \\
\textbf{ScanTextGAN} &0.984 &0.977 &\textbf{0.759} &0.693 & \textbf{0.931}\\
ScanTextGAN -- Text Reconstruction &0.986 &0.981 &0.756 &\textbf{0.706} & 0.939\\
% *ScanTextGAN + Dwell Time reconstruction & 0.985 &0.977	&\textbf{0.773}	&0.605 \\
% *Dwell Time reconstruction w/o Adv. training &0.990	&0.985 &0.686	&0.701 \\
ScanTextGAN -- GAN Loss &0.990 &0.984 &0.739 &\textbf{0.706} &0.945\\
ScanTextGAN + addition of noise &0.984 &0.976 &0.759 &0.703 & 0.943\\
ScanTextGAN -- Text (CLS) Reconstruction + sentence reconstruction &0.983 &0.974 &0.667 &0.674 & 0.958\\
\bottomrule
\end{tabular}}
\caption{Cross-domain Evaluation of Scanpath Generation on the Dataset by \citet{Mishra_Kanojia_Nagar_Dey_Bhattacharyya_2017}.}\label{tab:iitb_2017}
%\vspace*{-3mm}
\end{table*}

\textbf{Results:} Table~\ref{tab:celer} presents the results of our scanpath prediction model on the CELER dataset. Further, we also compare ScanTextGAN with baselines on two other contemporary datasets of movie reviews, tweets, and sarcastic quotes \cite{mishra2016predicting},  Wikipedia and simple Wikipedia paragraphs  \cite{Mishra_Kanojia_Nagar_Dey_Bhattacharyya_2017}. Tables~\ref{tab:iitb_2016} and \ref{tab:iitb_2017} present the results of our model on those datasets. For obtaining results on these corpora, we use the model trained on the CELER dataset, thus helping us evaluate the cross-domain performance of the model. 

As can be seen in Table~\ref{tab:celer}, Table~\ref{tab:iitb_2016} and Table~\ref{tab:iitb_2017}, ScanTextGAN outperforms other models for scanpath prediction on most metrics. The performance of ScanTextGAN even surpasses inter-subject reference on Duration and comes very close to Vector, Length, and Position. 

We observe that adopting the reconstruction of the CLS token as an auxiliary task (Task - 2) boosts the model performance. Reconstructing the full sentence embeddings rather than the CLS tokens only as an auxiliary task does not always improve the results, despite adding a larger computational overhead. The results also reveal that concatenating noise with text embeddings is more rewarding than adding it.

Further, to compare the skipping behavior of ScanTextGAN with humans, we calculate the weighted F1 score of the words skipped and attended by both model types. We find the weighted F1 to be 64.6 between them. Fig.~\ref{fig:scanpath_comparison} presents a visual comparison between real scanpaths from the available eyetracking data and scanpaths generated by ScanTextGAN, corresponding to some randomly chosen text samples. We can observe that the generated scanpaths resemble the real ones to a great extent. Thus, the quantitative and qualitative results on in-domain and cross-domain settings lead us to believe that our proposed scanpath generation model can be deemed a good approximator of the human scanpaths. % and can be used for downstream NLP applications.

%%%%%%%%%%%%%%%%%%%%%%%%%%%%%%%%%%%%%%%%%%%%%%%%%%%%%%%%%%%%%%%%%%
%%%%%%%%%%%%%%%%%%%%%%%%%%%%%%%%%%%%%%%%%%%%%%%%%%%%%%%%%%%%%%%%%%
\subsection{Application to NLP Tasks}
\label{sec:Application to NLP Tasks}
% \begin{table}[t]\centering
% \caption{Sentiment analysis and sarcasm detection results on the dataset by \citet{mishra2016predicting}. Model configuration refers to the type of scanpath included in train and test data.}
% \label{tab:iitb_classifier_results}
% % \scriptsize
% \begin{tabular}{cc|ccc}\toprule
% \multicolumn{2}{c}{\textbf{Model Configuration}} &\multicolumn{2}{c}{\textbf{Weighted F1 score}} \\\cmidrule{1-4}
% \textbf{Train} & \textbf{Test} &\textbf{Sentiment} &\textbf{Sarcasm} \\\midrule
% None &None &0.7839 &0.9438 \\
% Real &Real &0.8334 &0.9501 \\
% Random &Generated &0.7773 &0.9313 \\
% Real &Generated &0.8319 &0.9378 \\
% Generated &Real &0.8402 &0.9452 \\
% Generated &Generated &0.8332 &0.9506 \\
% Real+Generated &Generated &\textbf{0.8404} &\textbf{0.9512} \\
% \bottomrule
% \end{tabular}
% \end{table}

% \usepackage{booktabs}

We use them to augment various NLP models and measure their performance to demonstrate the usefulness of cognitive signals hidden in the \textit{generated} scanpaths.

\textbf{Sentiment Classification and Sarcasm Detection:} For these tasks, we use a model consisting of a network of two branches of BiLSTMs and Batch Normalization layers that perform sequential modeling over text representations obtained through BERT and scanpaths fed as input to the model. The outputs of both branches are combined and passed to another layer of BiLSTMs, followed by a feed-forward network that predicts binary sentiment/sarcasm labels corresponding to the input after activating with the Sigmoid function. We follow a 10-fold cross-validation regime.

We compare the models with generated scanpaths, real scanpaths, and without scanpaths. Further, to investigate whether performance gains observed by adding scanpaths are due to scanpaths and not the increase in the number of parameters, we train a \textit{Random-Random} variant in which we send Random noise as scanpaths to the model with an increased number of parameters. We also simulate the real-world case where both real and generated scanpaths are available during train time, but only generated ones are available during test time, for example, during user deployment. 
%We also compare the results for all tasks with models fed with random noise as scanpaths. 
\begin{table}[!t]
\centering
% \vspace{-10mm}
\resizebox{0.85\columnwidth}{!}{\begin{tabular}{llcc} 
\toprule
\multicolumn{2}{c}{\textbf{Model Configuration}} & \multicolumn{2}{c}{\textbf{F1 score}}  \\ 
\midrule
\textbf{Train}   & \textbf{Test}                 & \textbf{Sentiment} & \textbf{Sarcasm}           \\ 
\midrule
w/o              & w/o                           & 0.7839             & 0.9438                     \\
Random           & Random                     & 0.7990             & 0.9397    \\
Random           & Generated                     & 0.7773             & 0.9313                     \\
Real             & Generated                     & 0.8319             & 0.9378                     \\
Real      & Real        & 0.8334        & 0.9501             \\
Generated        & Real                          & 0.8402             & 0.9452                     \\
Generated        & Generated                     & 0.8332             & 0.9506                     \\
Real + Generated & Generated                     & \textbf{0.8404}    & \textbf{0.9512}            \\ \midrule

Intent-Aware & Intent-Aware & \textbf{0.8477} & \textbf{0.9528} \\ \bottomrule
\end{tabular}}
\caption{Sentiment analysis and sarcasm detection results on the dataset by \citet{mishra2016predicting}. Model configuration refers to the type of scanpath included in train and test data.\label{tab:iitb_classifier_results}
}
%\vspace{-3mm}
\end{table}

Table~\ref{tab:iitb_classifier_results} records the results of sentiment analysis and sarcasm detection tasks \cite{mishra2016predicting}. We note that generated scanpaths training and testing lead to similar gains for sentiment analysis and sarcasm detection as real scanpaths. The model with an increased number of parameters fed random noise in place of scanpaths performs similarly to the model trained without any scanpaths. Interestingly, the best results are obtained when model training uses both real and generated scanpaths. We believe this is due to ScanTextGAN bringing additional cognitive information from the news-reading CELER corpus, which is not present in the real scanpaths in \citet{mishra2016predicting}. In addition to the intrinsic evaluation presented in \S\ref{sec:Evaluation of Scanpath Generation}, this downstream evaluation demonstrates the high quality of the synthesized scanpaths, showing that they contain valuable cognitive processing signals for NLP tasks.

\textbf{GLUE Tasks}:
To validate further, we augment classification models (based on sequential modeling using LSTMs) with generated scanpaths to show performance improvement in downstream NLP tasks on four GLUE benchmark datasets – SST, MRPC, RTE, QQP as described in \S\ref{sec:eval_datasets}.
% (Since our goal is to establish the hypothesis that scanpaths lead to performance improvement in NLP models and do not necessarily achieve state-of-the-art results, we use simple classification models.)
Table~\ref{tab:glue_results} reports the accuracy and weighted-F1 scores of the models trained with and without scanpaths for these tasks. We observe that in all four tasks, the model trained with generated scanpaths outperforms the one without scanpaths.

% \begin{table}[!htp]\centering
% %\vspace{-2mm}
% \caption{\small Results on GLUE benchmark tasks.}\label{tab:glue_results}
% % \scriptsize
% \begin{tabular}{lccc}\toprule
% \textbf{Model} &\textbf{Acc} &\textbf{F1} \\\midrule
% \textit{Dataset} &\multicolumn{2}{c}{\textbf{SST}} \\\midrule
% Without scanpaths &0.8090 &0.8089 \\
% With generated scanpaths &\textbf{0.8138} &\textbf{0.8138} \\\midrule
% \textit{Dataset} &\multicolumn{2}{c}{\textbf{MRPC}} \\\midrule
% Without scanpaths &0.6902 &0.6656 \\
% With generated scanpaths &\textbf{0.6969} &\textbf{0.6828} \\\midrule
% \textit{Dataset} &\multicolumn{2}{c}{\textbf{RTE}} \\\midrule
% Without scanpaths &0.6162 &0.6080 \\
% With generated scanpaths &\textbf{0.6211} &\textbf{0.6205} \\
% \bottomrule
% \end{tabular}
% %\vspace{-2mm}
% \end{table}

\begin{table}[!t]
\centering
% \vspace{-10mm}
\resizebox{0.85\columnwidth}{!}{\begin{tabular}{llcc} 
\toprule
\textbf{Dataset}      & \textbf{Model}         & \textbf{Acc}    & \textbf{F1 score}  \\ 
\midrule
\multirow{2}{*}{SST}  & w/o scanpaths           & 0.8090           & 0.8089             \\
                      & w/ random scanpaths & 0.8059 & 0.8061  \\
                      
                      & w/ generated scanpaths & \textbf{0.8138} & \textbf{0.8138}  \\\cmidrule{2-4}
                      & w/ intent-aware scanpaths & \textbf{0.8269} &	\textbf{0.8272}
                      \\\midrule
\multirow{2}{*}{MRPC} & w/o scanpaths           & 0.6902          & 0.6656             \\
                      & w/ random scanpaths & 0.6623 & 0.6680  \\
                      & w/ generated scanpaths & \textbf{0.6969} & \textbf{0.6828}   \\\cmidrule{2-4}
                      & w/ intent-aware scanpaths & \textbf{0.7009} &	\textbf{0.6911}
                      \\\midrule
\multirow{2}{*}{RTE}  & w/o scanpaths           & 0.6162          & 0.6080              \\
                      & w/ random scanpaths & 0.5802 & 0.5794  \\
                      & w/ generated scanpaths & \textbf{0.6211} & \textbf{0.6205}   \\\cmidrule{2-4}
                      & w/ intent-aware scanpaths & \textbf{0.6293} &	\textbf{0.6278}
                      \\\midrule
\multirow{2}{*}{QQP}  & w/o scanpaths           & 0.8499          & 0.8513              \\
                    & w/ random scanpaths & 0.8491 & 0.8503  \\
                    & w/ generated scanpaths & \textbf{0.8578} & \textbf{0.8596}    \\\cmidrule{2-4}
                      & w/ intent-aware scanpaths & \textbf{0.8648} &	\textbf{0.8658} \\
\bottomrule
\end{tabular}}
\caption{Results of training NLP models with and without scanpaths on the GLUE benchmark tasks. Including scanpaths leads to consistent improvements across all the NLP tasks.\label{tab:glue_results}}
%\vspace{-3mm}
\end{table}

\iffalse
    \begin{table*}[!h]\centering
    \caption{Results of Sentiment Analysis and Sarcasm Detection tasks on \citet{mishra2016predicting}}
    \label{tab:iitb_classifier_results}
    % \scriptsize
    \begin{tabular}{cccccccc}\toprule
    \multicolumn{5}{c}{\textbf{Model Configuration}} &\multicolumn{2}{c}{\textbf{Weighted F1 score}} \\\cmidrule{1-7}
    \multicolumn{3}{c}{\textbf{Train}} &\multicolumn{2}{c}{\textbf{Test}} &\multirow{2}{*}{\textbf{Sentiment Analysis}} &\multirow{2}{*}{\textbf{Sarcasm Detection}} \\\cmidrule{1-5}
    Generated &Real &Random &Generated &Real & & \\\midrule
    \xmark &\xmark &\xmark &\xmark &\xmark &0.7839 &0.9438 \\
    \xmark &\xmark &\cmark &\cmark &\xmark &0.7773 &0.9313 \\
    \cmark &\xmark &\xmark &\cmark &\xmark &0.8332 &0.9506 \\
    \xmark &\cmark &\xmark &\xmark &\cmark &0.8334 &0.9501 \\
    \xmark &\cmark &\xmark &\cmark &\xmark &0.8319 &0.9378 \\
    \cmark &\cmark &\xmark &\cmark &\xmark &\textbf{0.8404} &\textbf{0.9512} \\
    \bottomrule
    \end{tabular}
    \end{table*}
    
\fi
 
\textbf{Intent-Aware Scanpaths:} \label{sec:intent-scanpaths} Finally, we try to condition scanpaths generation on the downstream natural language task. We back-propagate gradients from the downstream NLP task to the conditional generator. In this fashion, the model learns to generate \textit{intent-aware} scanpaths.
The hypothesis is that finetuning scanpath generation based on feedback from the natural language task will bias the generator towards words more pertinent to that task and thus could help further improve performance on the downstream task. The architecture is shown in Appendix: Fig~\ref{fig:intent-model}. The results in Tables~\ref{tab:iitb_classifier_results} and \ref{tab:glue_results} validate the hypothesis that we observe consistent improvements in all downstream tasks. Fig~\ref{fig:intent-scanpaths-example} and Appendix: Fig~\ref{fig:intent-saliency-example} show a few examples of scanpaths and saliency generated for three downstream natural language tasks. 

Together these results corroborate the hypothesis that leveraging the cognitive signals approximated by synthetic scanpaths in NLP models leads to performance gains.

\section{Conclusion}
\label{sec:ConclusionFutureWork}
In this work, we make two novel contributions toward integrating cognitive and natural language processing. (1) We introduce the first scanpath generation model over text, integrating a cognitive reading model with a data-driven approach to address the scarcity of human gaze data on text. (2)~We propose generated scanpaths that can be flexibly adapted to different NLP tasks without needing task-specific ground truth human gaze data. We show that both advances significantly improve performance across six NLP datasets over various baselines. Our findings demonstrate the feasibility and significant potential of combining cognitive and data-driven models for NLP tasks. Without the need for real-time gaze recordings, the potential research avenues for augmenting and understanding NLP models through the cognitive processing information encoded in synthesized scanpaths are multiplied.

\section{Limitations}
\label{Limitations}
In this work, we demonstrated artificial scanpath generation over multiple eye-tracking datasets. Further, our experiments build a link between cognitive and natural language processing and show how one can inform the other. However, the proposed method has a few limitations, which we aim to address in the future. The field needs work on bigger and more diverse eye-tracking datasets, which can enable scanpath generation over longer text sequences and can model generating scanpaths conditioned on previously read context. Besides, a better understanding of the entire scanpath generation process can help model the intra and inter-sentence scanpath generation process. The understanding would enable the integration of scanpaths to generative modeling tasks, which we intend to take up in future work. Another parallel direction is to include both explicit (like using RLHF) and implicit signals (like using cognitive signals) to better NLP tasks like language modeling.

% \clearpage
\bibliographystyle{acl_natbib}
\bibliography{anthology,custom_eacl2023}

% Entries for the entire Anthology, followed by custom entries
\clearpage
% \vfill\null
% \columnbreak

\appendix

\section{Scanpath Evaluation Metrics}
\label{sec:appendix_scanpath_metrics}
\looseness=-1
\textbf{MultiMatch} is a geometrical measure that models scanpaths as vectors in 2-D space, wherein the vectors represent saccadic eye movements. Starting and ending coordinates of these saccades constitute the fixation positions. It compares scanpaths across multiple dimensions, \textit{viz.} shape, length, position, direction, and fixation duration. Shape measures the vector difference between aligned saccade pairs, which is then normalized by twice the diagonal screen size. Length measures the normalized difference between the endpoints of real and generated saccade vectors. Direction is the angular distance between the two vectors. The position is the Euclidean difference in position between aligned vectors, and duration measures the difference in fixation durations normalized against the maximum duration. Since our work deals with scanpaths over text, we use 1-D space to represent the saccade vectors where word IDs denote the fixation positions. Thus, it is easy to see that computing scanpath direction similarity is redundant here (it is subsumed within position); hence we drop it from our analysis.

\textbf{Levenshtein Distance} between a pair of sequences measures the least number of character edits, i.e., insertion, deletion, and substitution needed to transform one sequence into the other.
Specifically, we use it to gauge the degree of dissimilarity between a pair of real $R$ and generated $G$ scanpaths. To account for the fixation durations of each word, $R$ and $G$ are temporally binned using a $50$ ms bin size, similar to the computation of ScanMatch metric \cite{cristino2010scanmatch}. The resulting sequences of word IDs, $R_W$ and $G_W$ are transformed into character strings, $R_S = \{r_1, r_2, ..., r_n\}$ and $G_S = \{g_1, g_2, ...,g_m\}$, where $R_S$ and $G_S$ are strings over the ASCII alphabet and $n = |R_S|$ and $m = |G_S|$.

Levenshtein Distance (LD) between strings $R_S$ and $G_S$ is computed and then normalized by the length of the longer string, which yields a Normalized Levenshtein Distance (NLD) score, as given below:
\begin{equation}
    NLD = \frac{LD(G_S, R_S)}{\max( |R_S|, |G_S| )}
\end{equation}
Thus, a lower NLD score is indicative of greater scanpath similarity.

\section{Intent-Aware Scanpaths}
\label{sec:appendix_intent_aware_scanpaths}
\begin{figure*}[!t]
%\vspace*{-1in}
    \centering
    \includegraphics[scale=0.55]{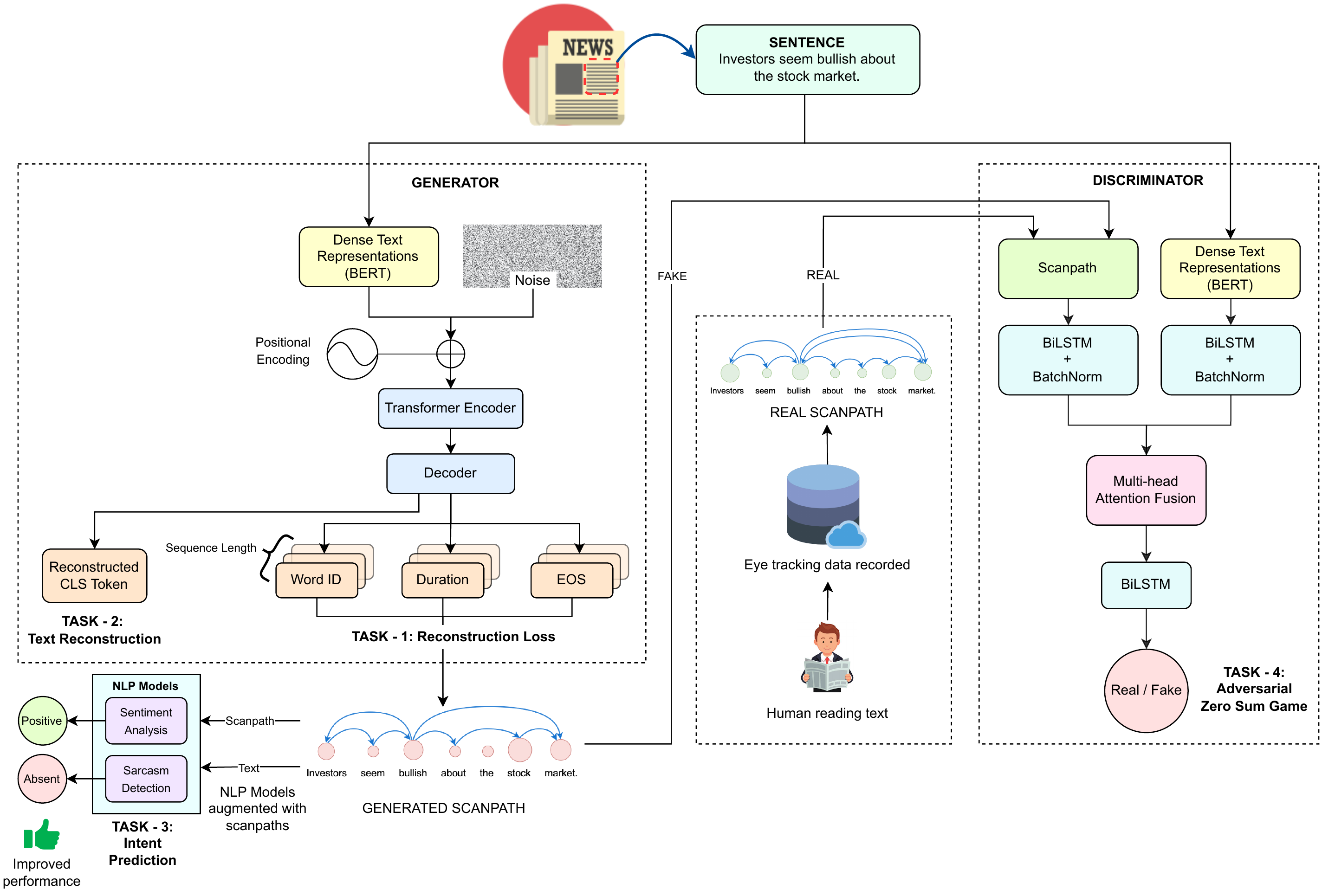}
    %\vspace*{-2mm}
    \caption{The architecture of the proposed Intent-Aware \textbf{ScanTextGAN} model. The model consists of a conditional generator and a discriminator playing a zero-sum game. Two cognitively inspired losses train the generator: scanpath (Task-1) and text (Task-2) reconstruction, a loss from the downstream intent of the natural language task (Task-3), and finally, the loss from the adversarial zero-sum game (Task-4). Variations of scanpaths are generated based on the downstream natural language task.}
    \label{fig:intent-model} 
%\vspace*{-5mm}
\end{figure*}

\begin{figure*}[]
%\vspace*{-2in}
    \centering
    \includegraphics[width=0.9\textwidth]{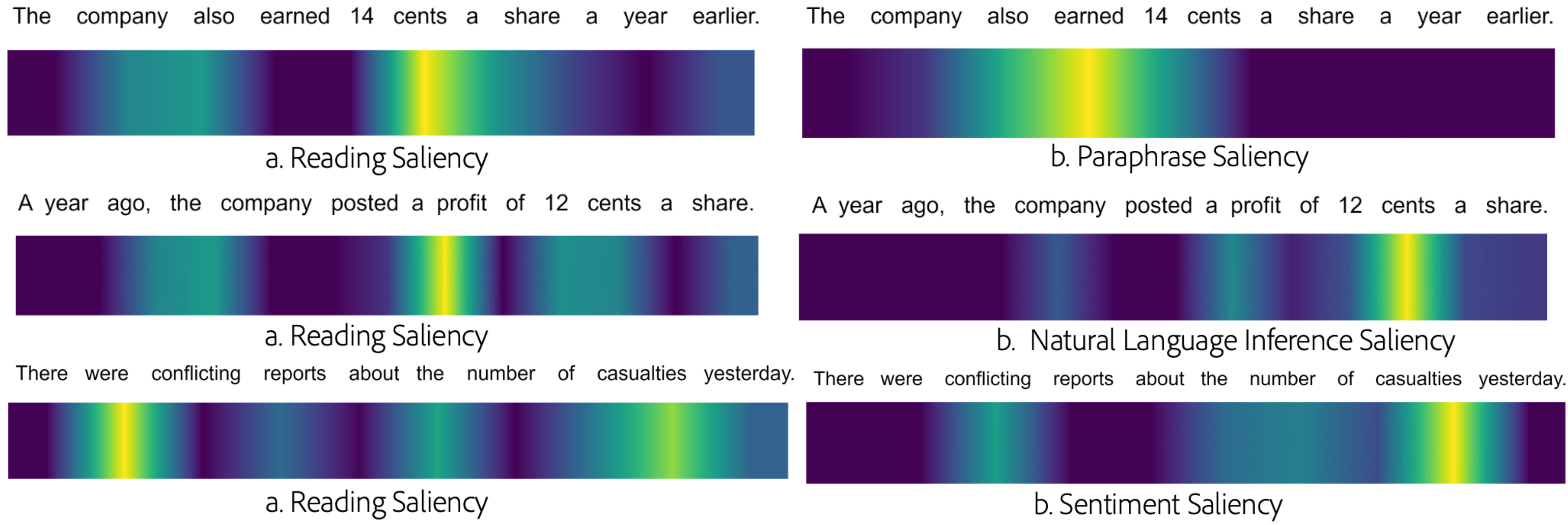}
    %\vspace*{-2mm}
    \caption{Saliency samples generated by conditioning scanpath generation on different downstream natural language tasks. It can be observed that the conditioned saliency pays much more attention to words important for that downstream task.}
    \label{fig:intent-saliency-example} 
%\vspace*{-5mm}
\end{figure*}

As described in section \S\ref{sec:Application to NLP Tasks}, the generator conditioned on the downstream natural language task yields \textit{intent-aware} scanpaths. Augmenting NLP models with these scanpaths leads to higher performance gains. Here, we provide more details on \textit{intent-aware} scanpath generation.
Please refer to figures \ref{fig:intent-model} and \ref{fig:intent-saliency-example} on the following page. Saliency corresponding to intent-aware scanpaths are shown in Fig.~\ref{fig:intent-saliency-example}.

\end{document}